\documentclass[epsf,prb,twocolumn,showpacs,preprintnumbers]{revtex4}
\usepackage{graphics}
\usepackage{graphicx}
\usepackage{dcolumn} 
\usepackage{bm}
\usepackage{color}
\usepackage{epsfig}
\begin{document}
\title{Structural and Correlation Effects in the Itinerant 
 Insulating Antiferromagnetic Perovskite NaOsO$_3$
}
\author{Myung-Chul Jung$^1$, Young-Joon Song$^1$}
\author{Kwan-Woo Lee$^{1,2}$}
\email{mckwan@korea.ac.kr}
\author{Warren E. Pickett$^{3}$}
\email{pickett@physics.ucdavis.edu}
\affiliation{
$^1$Department of Applied Physics, Graduate School, Korea University, Sejong 339-700, Korea\\
$^2$Department of Display and Semiconductor Physics, Korea University, Sejong 339-700, Korea\\
$^3$Department of Physics, University of California, Davis, California 95616, USA
}
\date{\today}
\pacs{71.20.Be, 71.30.+h, 71.27.+a, 75.50.Ee}
\begin{abstract}
The orthorhombic perovskite NaOsO$_3$ undergoes a continuous metal-insulator transition (MIT), 
accompanied by antiferromagnetic (AFM) order at $T_N=410$ K, suggested to be an example of
the rare Slater (itinerant) MIT.
We study this system using {\it ab initio} and related methods, focusing on the origin
and nature of magnetic ordering and the MIT.
The rotation and tilting of OsO$_6$ octahedra in the GdFeO$_3$ structure result in moderate narrowing
the band width of the $t_{2g}$ manifold, but sufficient to induce flattening of bands and AFM order
within the local spin density approximation (LSDA), where it
remains metallic but with a deep pseudogap.
Including on-site 
Coulomb repulsion $U$, at $U_{c}\approx 2$ eV
a MIT occurs only in the AFM state.
Effects of spin-orbit coupling (SOC) on the band structure seem minor 
as expected for a half-filled 
$t_{2g}^{3}$ shell, but SOC doubles
the critical value $U_{c}$ necessary to open a gap and also leads to large magnetocrystalline
energy differences in spite of normal orbital moments no greater than 0.1$\mu_B$.
Our results are consistent with a Slater MIT driven by magnetic order, induced by a 
combination of structurally-induced band narrowing and moderate Coulomb repulsion, with
SOC necessary for a full picture.
Strong $p-d$ hybridization reduces the moment, and when bootstrapped by the reduced Hund's
rule coupling (proportional to the moment) gives a calculated moment of
$\sim$1 $\mu_B$, consistent with the 
observed moment and only a third of the formal $d^3$ value. We raise and discuss 
one important question:
since this AFM ordering is at $q$=0 (in the 20 atom cell) where nesting is a moot issue,
what is the microscopic driving force for ordering and the accompanying MIT?
\end{abstract}
\maketitle

\section{Introduction}
In condensed matter physics the origins of, and phenomena associated with,
metal-insulator and  magnetic transitions 
remain issues of fundamental interest, especially when they do not fit into a
conventional mold. From this viewpoint, a few $5d$-based oxides, including the
osmate studied here, that display various high formal oxidation states
are attracting great interest. They display an abundant variety of physical phenomena, including
superconductivity,  important effects due to large spin-orbit coupling (SOC), 
unconventional metal-insulator transitions (MIT) (the focus of this paper), 
and unusual magnetic behavior.

Within heptavalent Os$^{7+}$ $d^1$ systems, the cubic double perovskite Ba$_2$NaOsO$_6$ is ferromagnetic
with an unusually small magnetic moment while maintaining a cubic structure,\cite{bnoo} 
whereas the magnetic structures of the trigonal Na$_3$OsO$_5$, 
Li$_5$OsO$_6$, and the double perovskite Ba$_2$LiOsO$_6$ 
are based on antiferromagnetic (AFM) order.\cite{na3oso5,li5oso6,bloo}
In the hexavalent $d^2$ systems, the double perovskite Ba$_2$CaOsO$_6$ is antiferromagnetic.\cite{ba2caoso6}
At the metallic $d^{2.5+}$ oxidation level, KOs$_2$O$_6$ in the $\beta$-pyrochlore structure
shows superconductivity\cite{kos2o6} at $T_c$=8 K,
whereas the triple perovskite Ba$_3$LiOs$_2$O$_9$
becomes AFM\cite{tripero} at 13 K, but shows significant reduction in the effective moment
of 3.34 $\mu_B$ from the theoretical spin-only value of 4.78 $\mu_B$, suggesting
strong SOC effects.
In a pentavalent $d^3$ system, the slightly distorted double perovskite Sr$_2$CrOsO$_6$ 
is a semimetallic ferromagnet in which SOC plays an important part in the magnetic
behavior, by reducing the spin moment by 0.27 $\mu_B$ and inducing an orbital moment 
of 0.17 $\mu_B$ on the Os ion, decreasing the net moment by nearly a factor of two.\cite{scroo,LP08}
Pentavalent Cd$_2$Os$_2$O$_7$, with its magnetic frustration on the pyrochlore Os sublattice,
has been suggested to be an itinerant Slater insulator,\cite{slater} 
{\it i.e.} a  MIT induced by antiferromagnetic ordering 
rather than by strong correlation effects.\cite{mandrus,cd2os2o7} 
Regarding Slater insulating states, $d^5$ Sr$_2$IrO$_4$, with a 
weakly canted antiferromagnetic MIT\cite{crawford} at 250 K,
has recently been characterized as a Slater insulator.\cite{kunes}
The character of the non-collinear ordered $d^3$ Cd$_2$Os$_2$O$_7$  
is still under discussion.\cite{yamaura}

The orthorhombic perovskite NaOsO$_3$, also with pentavalent Os, synthesized by Shi {\it et al.}, undergoes a MIT coincident with antiferromagnetic ordering at the rather high 
temperature of $T_N$=410 K with no evidence for any further structural distortion below 600 K.\cite{shi} 
The resistivity increases, and the
Hall number decreases, by $\sim$3 orders of magnitude from T$_N$ to 200 K and continued 
decreasing to 2 K, the insulating character becoming more accelerated below 25 K
toward a strongly insulating ground state.
The electronic contribution to
the specific heat measurement vanished at low temperature, indicating a fully gapped
Fermi surface. 
Due to this uncommon behavior, seemingly a continuously increasing gapping of the
Fermi surface leading to decrease and finally vanishing of carriers, 
this system was argued to be the first example
of the continuous MIT (so-called the Slater insulator with continuous opening
of the AFM gap). 

Recently, Calder {\it et al.} confirmed the continuous transition as a
simultaneous change in magnetic and transport character.\cite{calder}
The effective Curie moment of 2.71 $\mu_B$ observed by Shi {\it et al.} 
is reasonably consistent with a $S=3/2$ moment, reduced (by 30\%) by hybridization with oxygen
and perhaps compensated somewhat by an orbital moment.
This picture is one of high spin $d^3$ Os$^{5+}$.[\onlinecite{shi}]
Neutron power diffraction (NPD) measurements indicate G-type antiferromagnetism
(G-AFM) with direction of moments alternating on neighboring Os ions,
but an Os {\it ordered moment} of only 1 $\mu_B$.\cite{calder} This reduced value is 
difficult to reconcile with a $S=3/2$ picture, since three dimensional arrays of
large spins should not be subject to reduction by fluctuations.
In the only previous theoretical work on NaOsO$_3$, 
Du {et al.} calculated a similar (even somewhat smaller) value from the sort
of first principles calculations we will describe and use below.\cite{wan}
The NPD measurements show the spin oriented along the $c$-axis 
without canting of the spins,\cite{calder}
which provides additional information related to the impact of large SOC,
one aspect that we want to investigate more thoroughly.

Du {\it et al.} obtained a magnetic ground state of the observed G-AFM type,
and provided a comparison of this state to other types of magnetic order that they could
obtain in their calculations for NaOsO$_3$.\cite{wan}
There remain several unresolved questions: 
(1) what is the origin and the nature of the AFM ordering, and what other states are
energetically nearby? (2) why is the ordered
moment only 1 $\mu_B$ for pentavalent Os $d^3$?
is strong SOC a factor? (3) why is this MIT continuous instead of the
more common first-order type of MIT in transition metal oxides? 
(4) does SOC perhaps play a more central role
than is apparent from the results presented by Du {\it et al.}? The orientation of the moments 
(magnetocrystalline anisotropy) may provide useful
conditions to address this question.

For another perovskite-based osmate Ba$_2$NaOsO$_6$, 
we observed strong interplay between correlation
effects, small bandwidth, and strong SOC that are necessary to consider together
to build an interpretation of the unusual features of that system,\cite{bnoo} which include
that it is an uncommon {\it ferromagnetic} Mott insulator, it displays a very small ordered
moment, and it remains cubic when it seems the $d^1$ configuration should favor at
least a small Jahn-Teller distortion.
In this paper, we address most of the questions we raised
above, using 
first principles spin-polarized calculations 
including both correlation effects and SOC.

\begin{figure}[tbp]
{\resizebox{7cm}{7cm}{\includegraphics{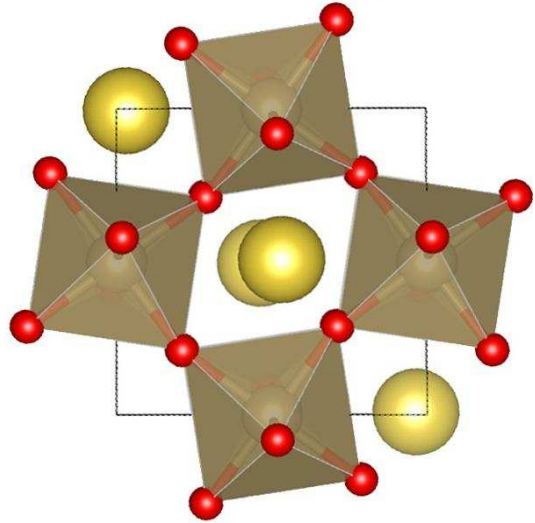}}}
\caption{(Color online) Distorted $Pnma$ crystal structure of NaOsO$_3$,
illustrating the $\sqrt{2}\times\sqrt{2}$ doubling cell in the $a$-$c$ plane.
There is also a doubling along the $b$ axis due to tilting.
The red (small) and the yellow (large) spheres indicate oxygen and Na atoms, 
respectively. This figure was produced by {\sc vesta}.\cite{vesta}
}
\label{str}
\end{figure}

\begin{table}[bt]
\caption{ Atomic positions of atoms of NaOsO$_3$ in the $Pnma$ (No. 62) structure,
 as experimentally observed.\cite{shi} 
}
\begin{center}
\begin{tabular}{ccccccc}\hline\hline
    &site symmetry  &  $x$  &~ & $y$ &~ & $z$    \\\hline
  Na  & $4c$ & 0.0328& & $\frac{1}{4}$ & &--0.0065\\
  Os  & $4b$ & 0    & & 0 & & $\frac{1}{2}$ \\
 apical O & $4c$ & 0.4834 & & $\frac{1}{4}$ & & 0.0808  \\
 planar O & $8d$ & 0.2881 & & 0.0394 & & 0.7112 \\ \hline
\end{tabular}
\end{center}
\label{table1}
\end{table}

\section{Structure and Methods}
Shi {\it et al.} reported a highly distorted structure involving 
relative displacement of Na and O ions, leading to a $\sqrt{2}\times 2\times \sqrt{2}$
quadrupled gadolinium orthoferrite type supercell (space group: $Pnma$, No. 62).\cite{shi}
This large structure distortion is consistent with the small tolerance factor
$t$=0.87, using the Shannon ionic radius.\cite{shannon}
This GdFeO$_3$ structure type has been described in some detail by Pickett and Singh.\cite{wep96}
The lattice parameters are $a$=5.3842, $b$=7.5804, and $c$=5.3282
(in units of \AA), corresponding to the volume per formula unit (f.u.) of a cube 
with edge 3.788 \AA. The atomic positions are given in table \ref{table1}.
This orthorhombic distortion leads to three somewhat different Os-O bond lengths, but
all are close to 1.94 \AA, resulting in a near-ideal  aspect ratio of $\sim$1.002.\cite{wep96} 
Thus the local structure is consistent with the $t_{2g}$ orbitals being equally
occupied ($t_{2g}^3$, S=$\frac{3}{2}$), retaining local cubic symmetry
in the OsO$_6$ octahedron. The O-Os-O bond angles are 90.7$^\circ$, 89.3$^\circ$, 
or 89.1$^\circ$, also indicating nearly ideal OsO$_6$ octahedra. 
However, the rotation and tilting of the octahedra are substantial.
The O-Os-O axis tilts by 11$^\circ$ relative to the $b$ axis and
rotates by 9$^\circ$ in the $a$ (or $c$) axis. 

In our work both SOC and correlation effects have been assessed, within LSDA+SOC, LDA+U,
and LSDA+U+SOC approaches, implemented in two all-electron full-potential codes {\sc fplo}\cite{fplo} 
and {\sc wien2k}.\cite{wien2k} (The latter combination is available only in {\sc wien2k}.)
To include correlation effects in the half-filled $t_{2g}$ shell, an effective on-site Coulomb repulsion
$U_{eff}=U-J$ was used for the results we present, 
since experience has shown results very similar 
to those obtained from separate $U$ and
the Hund's exchange integral $J$ inputs when a shell is half-filled. 
Convergence was checked up to a 14$\times$9$\times$14 $k$-mesh of Brillouin zone sampling.
In {\sc wien2k}, the basis size was determined by $R_{mt}K_{max}=7$, quite adequate 
for these atoms, and the APW radii
were Na 2.14, Os 2.15, and O 1.4, in {\it a.u.}.

\section{Magnetically ordered (T=0) phase}

\begin{figure}[tbp]
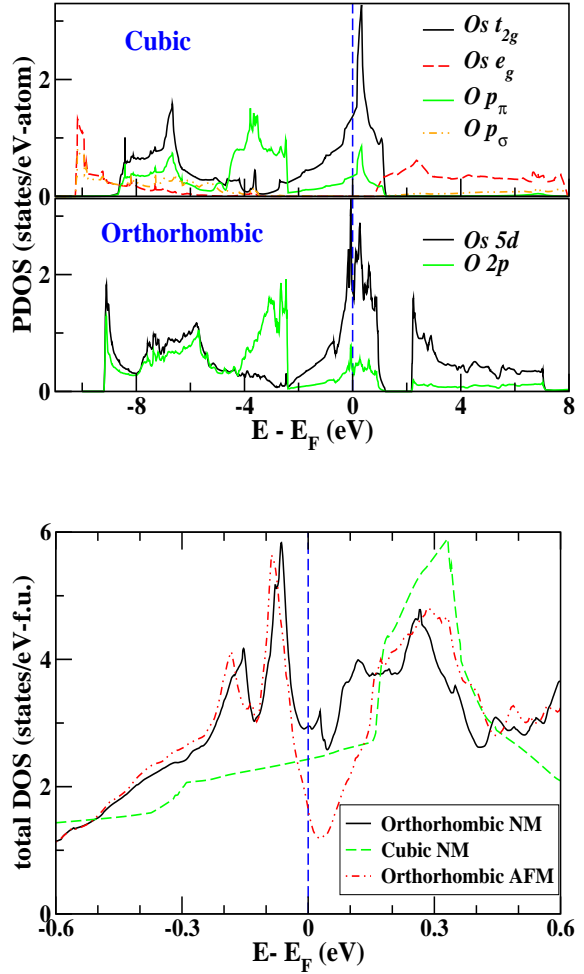

{\resizebox{7.5cm}{6.0cm}{\includegraphics{Fig2a.eps}}}
\vskip 9mm
{\resizebox{7.5cm}{6.0cm}{\includegraphics{Fig2b.eps}}}
\caption{(Color online) Top: orbital-projected densities of states (DOSs) for Os $5d$
 and O $2p$ in the cubic and distorted $Pnma$ crystal structures of nonmagnetic NaOsO$_3$.
The distorted phase DOS shows more structure around the Fermi energy $E_F$, 
reflecting more flat bands that enhance magnetic instabilities.
Bottom: enlarged total DOSs of the nonmagnetic cubic phase and the distorted (orthorhombic) 
nonmagnetic and G-AFM phases, near E$_F$, which is set to zero.
Notice that magnetic order strongly decreases N($E_F$).}
\label{pdos_nm}
\end{figure}

\begin{figure}[tbp]
{\resizebox{8cm}{7cm}{\includegraphics{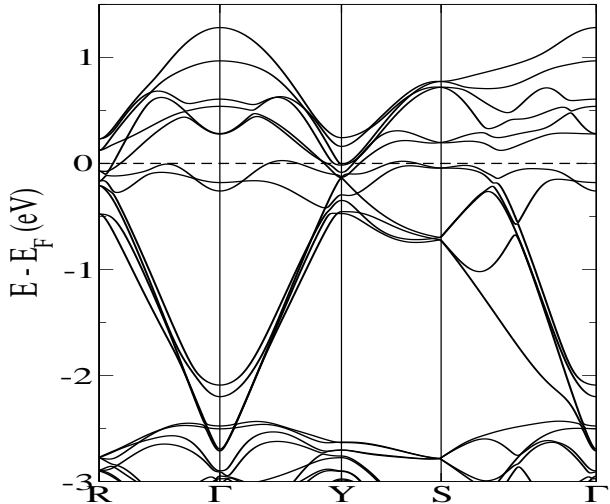}}}
\caption{AFM band structure in LSDA near $E_F$ 
(zero of energy), which is crossed by the Os $t_{2g}$ 
manifold. A deep and narrow pseudogap is centered at $E_F$. 
The Os $e_g$ bands lie higher, above the range of this figure, while
the O $2p$ bands lie below --2.5 eV. In units of $(\pi/a,\pi/b,\pi/c)$
the symmetry points shown are R=(111), $\Gamma$=(000), Y=(010), and
S=(110).
}
\label{band1}
\end{figure}

\subsection{LSDA electronic structure}
To begin, we compare electronic structures of the cubic and distorted phases.
Within LSDA, the cubic phase has a higher energy by 378 meV/f.u.
than the distorted structure in the nonmagnetic (NM) state, reflecting the
strong lowering of energy by the distortion alone. 
The atom-projected densities of states (DOSs) in both phases are compared 
in the top panel of Fig. \ref{pdos_nm} and are consistent with those presented
by Shi {\it et al.}\cite{shi}
As discussed in Ru-based perovskites by Mazin and Singh,\cite{singh97}
the OsO$_6$ octahedra must be understood as oxygen-shared clusters.
In the cubic phase, the $t_{2g}$-$p_\pi$ antibonding ($pd\pi^*$) and bonding ($pd\pi$) states 
lie roughly in the --2.9 to 1.3 eV and of --8.7 to --4.7 eV regions, respectively.
(The latter complex is typically neglected in characterizing the electronic
structure and in particular the oxidation level, leaving the $t_{2g}$-based 
$pd\pi^*$ half-filled as the physical $t_{2g}$ bands.)
Nearly pure $p_\pi$ bands extend from --5 to --2.5 eV.
The large orthorhombic distortion leads to narrowing bandwidth and enhancing 
the hybridization. The width of the $pd\pi^*$ manifold
is reduced by 15\%, and a hybridization gap of 1 eV appears 
between the $pd\pi^*$ manifold and the $e_{g}$-$p_\sigma$ ($pd\sigma^*$) antibonding manifold.

The bottom panel of Fig. \ref{pdos_nm} compares the total DOS N($E$) of
the NM cubic, NM distorted, and AFM distorted phases, near $E_F$.
In the cubic phase, N($E$) varies smoothly very near $E_F$ before giving
way to a large narrow peak centered 0.3 eV above $E_F$.
In the NM distorted phase, a sharp peak appears at --60 meV, and van Hove singularities appear
just above $E_F$ reflecting flat portions of bands that may encourage magnetic instability.
However, our fixed spin moment calculations show no evidence of any 
nearby stable or metastable ferromagnetic state.\cite{fsm}
In addition to sharp peaks on either side of $E_F$, the AFM DOS of the distorted phase shows 
a deep pseudogap just above E$_F$, suggesting that applying $U$ to the Os $5d$ states
at this band filling will open a gap.

Within LSDA, the AFM state with an Os spin moment of 0.6 $\mu_B$ 
has a {\it very slightly lower energy}
of 3 meV/f.u. than NM in the experimentally observed structure.
The local moment of oxygens is negligible due to their symmetric
placement between antialigned Os moments.
This degeneracy (within a fraction of $k_B T_N$/Os) rationalizes the existence of
a magnetic ordering transition in NaOsO$_3$.
Figure \ref{band1} presents the enlarged LSDA band structure,
showing only the Os $t_{2g}$ ({\it i.e.}, $pd\pi^*$) manifold with the width $W$ 
= 3.5 eV. The deep pseudogap in the DOS of Fig. \ref{pdos_nm} results from an
incipient gap formation arising as a combination of the distorted lattice and AFM order.
The valence (occupied) band complex has become nearly disjoint from the 
conduction (unoccupied) bands in nearly the entire zone, with slight overlap
remaining at the $Y$ and $R$ points.
The top two occupied bands are quite flat,
accounting for the sharp peaks at --0.1 and --0.2 eV 
as shown in the bottom panel of Fig. \ref{pdos_nm}.

\subsection{Conditions for antiferromagnetic ordering}
To probe the effects of the structural geometry on the AFM state, 
we varied both the volume and the angles of the OsO$_6$ octahedral rotation. 
In the cubic structure with the same volume as the experimentally observed structure,
no AFM order could be obtained.
To check further for magnetic order in the cubic phase, the lattice parameter of the cubic phase 
was varied in the range of $\pm$3\%.
No magnetic order was obtained, indicating that the AFM order is not induced by
simple variation of the Os-Os distance.
Also, as expected, no AFM order is induced by Na displacement.

Next we varied the two structural angles individually.
At a critical tilt of $\theta_c$=3$^\circ$, a hybridization gap between $pd\pi^*$ 
and $pd\sigma^*$ 
manifolds appears, accompanied by AFM ordering with an Os moment of 0.26 $\mu_B$.
Somewhat unexpectedly, this moment remains unchanged up to $\theta$=11$^\circ$, 
the observed tilt angle.
When rotating the octahedra by $\phi$, AFM with $M_{Os}$=0.13 $\mu_B$ 
appears at $\phi_c$=9$^\circ$,
where the overlap between the $pd\pi^*$ and $pd\sigma^*$ manifolds becomes very small.
In all cases the opening of a hybridization pseudogap around the Fermi level
accompanies the AFM order,
establishing a close connection in this compound between structure and magnetic order.

\begin{figure}[tbp]
{\resizebox{8cm}{3.6cm}{\includegraphics{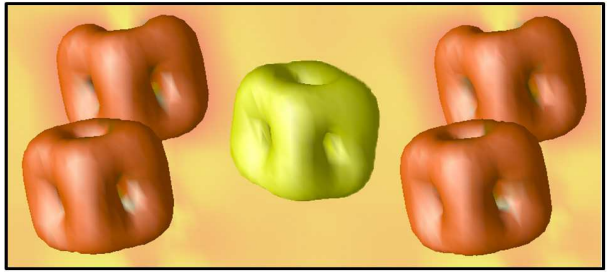}}}
\caption{(Color online)  Spin density plot of Os ions in the $a-c$ plane, 
with isosurface at 0.13 $e$/\AA$^3$.
This is obtained in LDA+U at $U_{eff}=1$ eV, where this system becomes
insulating, and reflects locally cubic symmetry.
The different color indicates the opposite spin orientation in the G-AFM alignment.
}
\label{spinden}
\end{figure}

\subsection{Metal-insulator transition versus U$_{eff}$}
Now we focus more closely on correlation effects beyond LSDA
on the AFM state in the observed distorted structure,
by applying the on-site Coulomb repulsion $U_{eff}$ to Os ions using the LDA+U method.
As might be anticipated from the pseudogap at $E_F$ at the required band filling,
a gap opens for $U_{eff}$ as small as 1 eV.
The Os local moment is 1.0 $\mu_B$, consistent with the experimentally observed value.\cite{calder}
This value is nearly unchanged when SOC is included (see below).
Considering $5d^3$ Os$^{5+}$ ions, the naive expectation based on a localized
$5d$ orbital is that orbital and spin ordering
of $t_{2g}^{2\uparrow} t_{2g}^{1\downarrow}$ (corresponding to a highly unusual spin density)
would be required to obtain this
reduced value of moment.
In the insulating state, the spin density however assumes a local cubic
symmetry dimpled dice shape,
shown in Fig. \ref{spinden}, which confirms equal occupation of the $t_{2g}$ orbitals.

\begin{figure}[tbp]
{\resizebox{8cm}{5.5cm}{\includegraphics{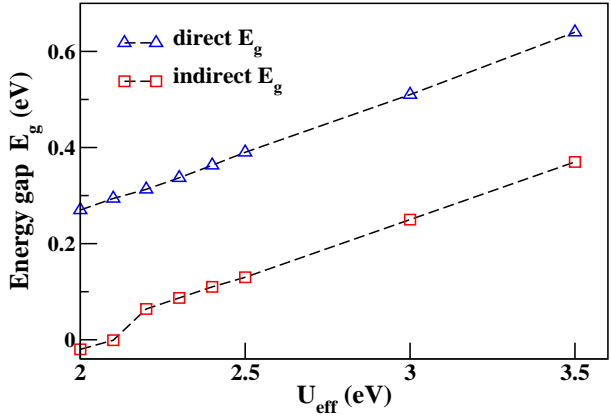}}}
\caption{(Color online) Direct and indirect gap vs. effective $U$ in LSDA+U+SOC
calculations. Below the critical value of $U^c_{eff}$=2.2 eV,
the indirect gap becomes negative, {\it i.e.,} the bands are semimetallic.
}
\label{gap}
\end{figure}

\begin{figure}[tbp]
\vskip 8mm
{\resizebox{8cm}{7cm}{\includegraphics{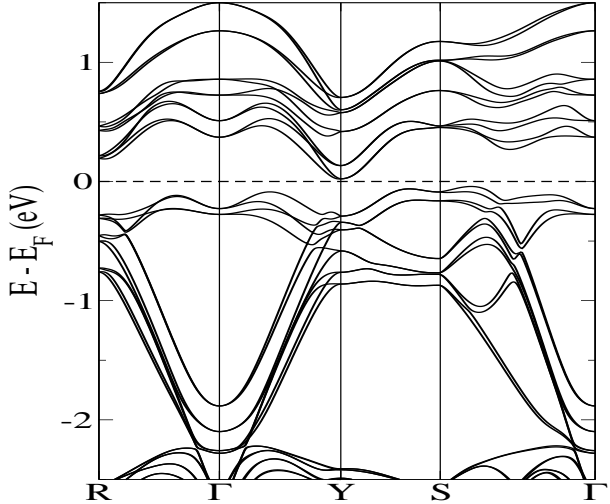}}}
\caption{Enlarged AFM band structure of Os $t_{2g}$ 
manifold at $U_{eff}$=2.2 eV in LSDA+U+SOC.
A gap opens at $U_{eff}$=1 eV in the LDA+U, but $U_{eff}$=2.2 eV is required when SOC is included.
}
\label{band2}
\end{figure}

\begin{figure}[tbp]
{\resizebox{8cm}{5.5cm}{\includegraphics{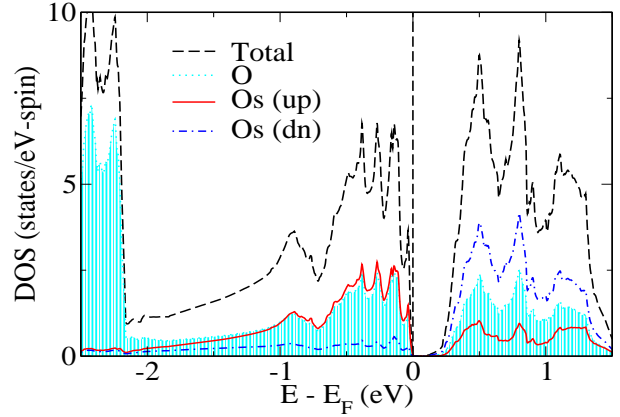}}}
\caption{(Color online) AFM total and atom-projected DOSs, 
including primarily the Os $t_{2g}$ manifold, within LSDA+U+SOC at $U^{c}_{eff}$=2.2 eV in the 
distorted structure.  In the occupied states below E$_F$, the unpolarized O $2p$
DOS mirrors closely the majority Os $t_{2g}$ DOS. The result is a 
greatly reduced spin moment on the Os ion (see text). 
}
\label{afmdos}
\end{figure}

\begin{table}[bt]
\caption{Local moments of Os in the distorted AFM NaOsO$_3$ in LSDA+U+SOC
calculations with the $\langle001\rangle$ quantization axis. 
The moments of oxygens are at most 0.01 $\mu_B$/atom.
}
\begin{center}
\begin{tabular}{ccccc}\hline\hline
 ~$U_{eff}$~ (eV) & state     & ~spin~  & ~~ &~orbital~     \\\hline
 1 & metallic & 0.69 & ~~ & --0.050 \\
 2 & semimetallic & 1.03 & ~~ & --0.085 \\ 
 2.5 & insulating& 1.13 & ~~ & --0.097 \\
 3.0 & insulating& 1.22 & ~~ & --0.106 \\\hline
\end{tabular}
\end{center}
\label{table2}
\end{table}

At $U_{eff}^c$, the $d$ occupation matrix substantiates this conclusion.
The majority $t_{2g}$ orbitals are almost equally occupied at about 0.7$e$
per orbital, substantially reduced from the formal value of unity. 
The majority $e_g$ and all the minority $d$ orbitals, which are formally unoccupied,
each have occupation of about 0.3$e$ per orbital, reflecting substantial $p-d$ hybridization
and indicating a strong deviation from the conventional ionic $d^3$ viewpoint.
In total there are 4.3 electrons throughout the $d$ orbitals, instead of the formal value 3.
This `discrepancy' is not so surprising, as it is widely recognized that an actual charge on
$d$ cations differs from the formal charge, 
sometimes\cite{wep12} by a factor of 2.  For Os here, the formal charge of +5 is contrasted
with an actual charge of around +3.7.  The reduction by a factor of three of the formal
moment is a more notable feature, and unlike the case of Ba$_2$NaOsO$_6$ where the moment is also
vastly reduced,\cite{LP08} here it is unrelated to SOC. 
NaOsO$_3$ is a case of an insulating transition metal
oxide where the formal charge (oxidation state) picture is murky at best, consistent with
the itineracy evident in the electronic structure and apparently in the MIT as well.

Figure \ref{gap} shows change in the direct and indirect gaps due to varying $U_{eff}$.
In the distorted structure and including SOC, an energy gap opens
only at $U_{eff}^c$=2.2 eV.[\onlinecite{Uc}]
Above $U_{eff}^c$, both gaps increase linearly at a rate of 0.16 eV/eV, 
with the difference of 0.3 eV independent of $U_{eff}$.
Below $U_{eff}^c$, a negative indirect gap (not shown) results due to significant lowering of the bottom
of the conduction band at the $Y$ point, though the direct gap is already established 
(the occupied and unoccupied bands are disjoint).
This feature may be related with the proposal of a continuous transition in opening an energy gap, 
as discussed in another osmate Cd$_2$Os$_2$O$_7$ by Shinaoka {\it et al.}\cite{cd2os2o7}  
Considering a likely $J\approx$0.5 eV, $U^c\approx$ 2.7 eV $\sim W$ suggests for this
multiband system that a Mott transition
is not the appropriate picture for this continuous MIT. 

The corresponding band structure at $U_{eff}^c$ is displayed in Fig. \ref{band2}.
The SOC splitting is about 0.2 eV at the $\Gamma$-point. 
An unusual outcome of the structural distortion is that the top of the valence bands is 
nearly flat, reflected in the step
increase in the valence band DOS evident in Fig. \ref{afmdos}, and providing a two-dimensional
character to the low energy phase space. 
We also point out that the dispersion is very similar above and below $E_F$,
relevant information for yet-to-be-done optical experiments since the $k$-conserving joint DOS may
be sharply peaked at the onset of absorption. 
Within LSDA+U+SOC as in LSDA, the AFM state is nearly degenerate with the NM state. 
The fact that neither state is energetically favored is consistent with the finite 
temperature transition between the
two states.

The value of $U$ appropriate for NaOsO$_3$ has already been mentioned in the literature,
in addition to the conclusion $U\sim$2 eV of Du {\it et al.} that is nearly identical
to ours. Calder {\it et al.} reported data on magnetic x-ray resonant scattering (MXRS)
with interpretation from x-ray absorption near edge spectra (XANES) obtained from
Joly's finite difference code.\cite{joly} They obtained a better correspondence using
$U=0$ than with nonzero $U$. We think there is no direct contradiction. In their
work they are dealing with resonant transitions between deep $2p_{1/2}$ and $2p_{3/2}$
levels and unoccupied $5d$ states. In our work, based on the self-consistent LDA+U
method, the effect of $U$ is to increase the separation of occupied $5d$ bands and
unoccupied $5d$ bands, which is a different issue involving a $U$ in a different 
manner.

\subsection{Additional effects of spin-orbit coupling}
As discussed for several decades, $t_{2g}$ orbitals when split off from the $e_g$
orbitals by the crystal field, form a representation of
an angular momentum of $L=1$.\cite{bnoo} In the spherical Hund's rule limit
the total angular momentum is zero for a high spin $t_{2g}^3$ system.
Especially in this locally-cubic geometry and half filling, the orbital
moment should be small.
As shown in Table \ref{table2}, inclusion of SOC leads to an orbital moment
antialigned with the spin moment and with magnitude no larger than
0.1 $\mu_B$.
Thus in the insulating state the net moment remains about 1 $\mu_B$,
consistent with the experimentally observed value,\cite{calder} 
and relatively insensitive to $U_{eff}$ in the range studied here (see Table \ref{table2}).

We have calculated total energies with LDA+U+SOC with four spin orientations
$\langle100\rangle$, $\langle010\rangle$, $\langle001\rangle$, and $\langle111\rangle$
to determine the easy axis.
As observed by Calder {\it et al.},\cite{calder} $\langle001\rangle$ is the easy axis.
Along $\langle010\rangle$, both spin and orbital moments increase by 10\% in magnitude,
but the energy cost is large, 474 meV/f.u., despite some additional Hund's coupling energy.
Along $\langle111\rangle$, the spin moment is enhanced to 1.25 $\mu_B$ while the energy 
is higher by 51 meV/f.u. 
Consistent with the similarities between the $a$ and $c$ axes (the octahedron rotation
is around the $b$ axis)
the energy difference is small, only 5 meV/f.u.


\section{Finite temperature; metal-insulator transition}
Approached from high temperature, a continuous MIT occurs at the point where the 
spin-symmetric phase becomes unstable 
to magnetic order, which is AFM order in the case of NaOsO$_3$.
In Slater-type (itinerant) transitions, ordering conventionally is driven by Fermi 
surface (FS) nesting, as in the classic case of chromium.
The common picture for a Slater MIT is that a spin density wave (SDW) gaps at least
a portion of the FS, and the increasing amplitude of the SDW destroys
more and more of the FS as the temperature is lowered.  The specific heat
data for NaOsO$_3$ indicate the FS is completely gone as T$\rightarrow$0. The
resistivity data strongly suggests it has been effectively destroyed at intermediate
temperature, since the magnitude and temperature coefficient reflect insulating
behavior even as spin-fluctuation scattering has began to be frozen out by the
magnetic order (which, in itself, should {\it lower} rather than increase the resistivity),
and the strongly decreasing Hall coefficient is consistent with rapidly shrinking FSs. 
In this section we first discuss the electronic structure in the (high temperature) 
nonmagnetic metallic state, 
albeit without consideration of the thermal broadening and the lattice dynamical and
spin fluctuation scattering that may be necessary for a detailed picture. Then we
provide a scenario for the MIT that is consistent with observations and what is
understood about the electronic structure.

\begin{figure}[tbp]
{\resizebox{7.5cm}{5.5cm}{\includegraphics{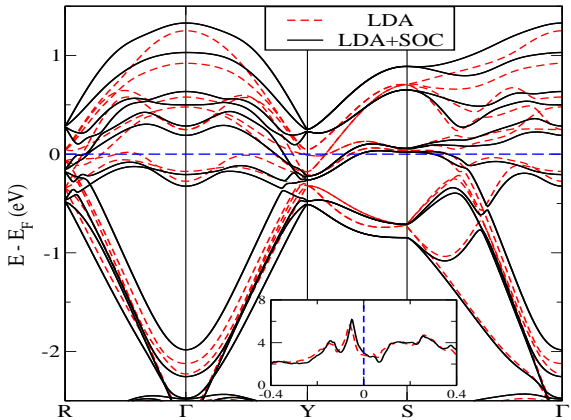}}}
\caption{(Color online) Enlarged view of the nonmagnetic $t_{2g}$ band structures
appropriate to the high T nonmagnetic phase, in LDA (red dashed lines)
and LDA+SOC (black solid line). Including $U$ makes negligible difference.
Inset: nonmagnetic total DOSs in LDA (red dashed lines) and LDA+SOC
(black solid lines) near $E_F$ (in states/eV-f.u.).
}
\label{nmband}
\end{figure}

\begin{figure}[tbp]
{\resizebox{7.5cm}{5.5cm}{\includegraphics{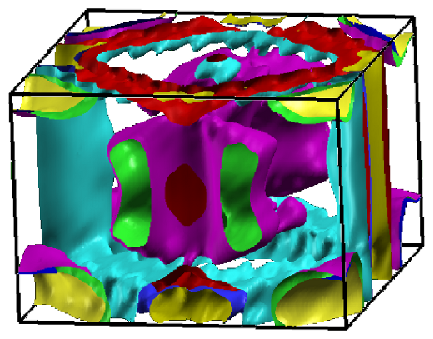}}}
{\resizebox{7.5cm}{5.5cm}{\includegraphics{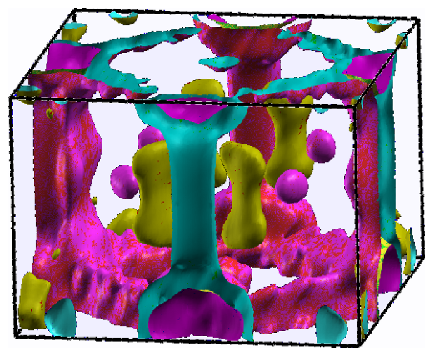}}}
\caption{(Color online) Fermi surfaces of LDA (top) and LDA+SOC (bottom)
in the nonmagnetic state. The $Y$ point is at the center of the front face.
In both cases complex multi-sheeted surfaces exist, but SOC results in
substantial rearrangement in Fermi surfaces.
}
\label{fs}
\end{figure}

\subsection{Electronic structure in the nonmagnetic phase}
The LDA and LDA+SOC band structures without magnetic order are displayed together in Fig. \ref{nmband}. 
We have checked that including U$_{eff}$ in this nonmagnetic phase leads to negligible
effect, as is commonly observed. In
the previous section we established that SOC has relatively little impact 
on the bands in the AFM
phase, displaying its main impact by its seeming competition with correlation effects so that the
correlation strength necessary to open the gap (within the AFM phase) is doubled to $\sim$2 eV.
Near $E_F$, the effect of SOC on the band structure (Fig. \ref{nmband}) in the high T phase again
seems minor, but it is clearly visible
at symmetry points (Y, S) where it removes degeneracies and splits bands. Moreover,
the effect on the DOS is minor, as is clear from the inset of Fig. \ref{nmband}. With or
without SOC, the Fermi level DOS N($E_F$) $\approx$ 3 states/eV-f.u. for both spins,
corresponding to a linear specific heat coefficient $\gamma$ = 7 mJ/mol-K$^2$. This
standard metallic value is corroborated by large FS sheets, which we now discuss. 

Because the topology of the FSs is central to a SDW transition, we display FSs without
and with SOC
in Fig. \ref{fs}.
The small shifts in band energies due to SOC that can be seen in Fig. \ref{nmband} 
are translated by the small
velocities into considerable changes in the complex multi-sheeted FSs, with some of the
clearest changes occurring around the symmetry points. The FSs retain to some extent an
approximate 4-fold symmetry in the $a$-$c$ plane ($x-y$ plane in Fig. \ref{fs}), which
of course is not a true symmetry of the $Pnma$ structure but is useful to recognize. 
There are both electron and hole type surfaces, making the magnitude of the Hall 
coefficient difficult to interpret.

\subsection{Comments on the metal-insulator transition}
Most theoretical work on AFM transitions in metals, including very recent ones, 
focus either on FS nesting (Cr is the classic example) or FS ``hot spots'' involving 
band and FS folding due to the finite-$q$ order that set in at the transition. In the
first case the nested sections of FS become gapped, while in the second case
the FS also reconstructs but in a less obvious manner. For 
example, Sachdev {\it et al.}\cite{sachdev} review the latter scenario and provide evidence
of a breakdown of fermionic quasiparticle excitations at the transition, and a
FS reconstruction from large to small surfaces that have been discussed for many
years in the high temperature superconducting cuprates.
  
Both the nesting scenario and the hot spot scenario of the continuous Slater 
transition to an AFM incurs difficulty
in NaOsO$_3$, because it is a {\it q=0 ordering} for which both scenarios, which
focus on a nonzero wavevector $Q_{AF}$, become moot.
This situation can arise, and not uncommonly does so, in crystal structures with more
than one magnetic ion of a given type. In NaOsO$_3$, the four symmetry-related Os ions
in the GdFeO$_3$ distorted lattice devolve into two pairs, with spins in opposite
directions.
Interestingly, the original exposition by Slater\cite{slater} did not involve finite-$q$ physics
in any explicit manner. His picture, based on spin-unrestricted Hartree-Fock theory, was
that AFM ordering would introduce new Fourier components in the (spin dependent) potential, leading
to new gaps in the band structure that could well appear at the Fermi level, especially
in stoichiometric compounds where there is the likelihood that bands can be completely
occupied, or completely unoccupied.  If the new gaps appear at a zone boundary, it will
be energetically favored to double the periodicity, as in the simplest picture of SDW order.
Yet if the new gaps appear only at $q$=0 (no cell doubling), the band 
structure may still become gapped in
spite of the lack of any extension of periodicity.

The effect of AFM ordering in NaOsO$_3$ on the electronic structure can be seen by
comparing the metallic bands of the nonmagnetic high T phase in Fig. \ref{nmband}
(solid lines) with the AFM insulating bands in Fig. \ref{band2}.  New Fourier components
of the potential separate the bands strongly at zone boundary symmetry points ({\it viz.}
R, Y, S) and at the zone center the bands are already split. Slater also tied this
type of ordering to bonding-antibonding band formation, which provides a band energy
(bonding energy) gain in addition to the Hund's rule (magnetic) gain in energy. 

A rudimentary picture that is consistent with data and with this understanding of the 
electronic structure can be assembled. Already at high T the strongly distorted, 
four f.u. per primitive cell, structure is dictated by the tolerance ratio of
0.87.  The distortion leaves flat bands at and near the FS, which is large and
multi-sheeted. At T$_N$, magnetic order sets in and the metallic band structure
begins to morph continuously toward the T=0 AFM insulating band structure. The
FSs, both electron and hole type, begin to shrink in volume and vanish one-by-one,
until no FS remains. The change from the metallic to the insulating band structure
should be monotonic in, and nearly proportional to, the magnetic order parameter.
The magnetic order parameter measured by Calder {\it et al.}
was provided only down to $\sim$0.85 T$_N$,\cite{calder} so it is not possible to estimate
when the last FS may be expected to disappear. As mentioned earlier, thermal smearing and
lattice and spin fluctuations will complicate a detailed picture. One loose
end in this picture is the apparent discontinuity -- a jump by roughly 50\% --
in the paramagnetic susceptibility\cite{shi} just below T$_N$.

Without a Fermi surface or hot spot origin, the mechanism of G-type ordering 
is not clear, and was not addressed by 
Du {\it et al.} in their theoretical study of magnetic ordering.\cite{wan}
Since G-type ordering
is common in perovskite transition metal oxides, it seems likely this ordering
represents an emergence of the nearest neighbor AFM coupling commonly occurring
in perovskites, in spite of the apparent
itineracy in the electronic structure and the collapse of the $d^3$ valence state
picture.

\section{Summary}
Our results have clarified a number of features of NaOsO$_3$, and the questions
posed in Sec. I. First, a combination
of the distortion of the perovskite structure and moderate correlation effects is
responsible for both antiferromagnetism and the insulating gap in the ground state.
Our calculations account naturally for the factor of three reduction in the observed
moment (1$\mu_B$) relative to the ideal $t_{2g}^3$ value, due to very strong
hybridization with the O $2p$ orbitals that render the conduction states, and the
magnetism, itinerant in nature.  The oxygen ions, being coordinated with two Os
ions with antialigned spins, are unpolarized by (approximate) symmetry.
The calculated energetics, with small energy difference between the magnetic
and non-magnetic states, is also consistent with
a Slater-type metal-to-insulator transition to an AFM state.
Moreover, the simplest specification of a Mott insulator is one in which the
insulating state does not depend on magnetic order. In NaOsO$_3$ all the evidence
indicates that magnetic order is {\it necessary} for insulating behavior.
Because of the itinerant nature of the moment, a Heisenberg model cannot
provide a representation which could then be applied to predict the Ne\'el temperature.

Spin-orbit coupling, often a dominant force in $5d$ metal oxides, has a more
modest effect in this compound.  While it effects the band structure enough that
the effective coupling strength U$_{eff}$ is increased from 1 eV to 2.2 eV, it is
in itself ineffective in decreasing the local moment.  The strong moment
reduction (from a localized high-spin $t_{2g}$ picture of 3$\mu_B$) 
is due to strong hybridization with O $2p$
orbitals -- that is, itineracy -- which is itself affected by the distortion. The minor importance of SOC 
is consistent with the finding of Calder {\it et al.} that xray resonant photoelectron
spectra show only minor effect of SOC.\cite{calder}

Though $3d$ perovskite oxides have been studied extensively for decades, their
$5d$ counterparts are relatively new to detailed inspection of their properties
and the underlying mechanisms. Our results for NaOsO$_3$ will help to provide
a basis for understanding the distinctions in the physics of $5d$ versus $3d$
(and $4d$) oxides.

\section{Acknowledgments}
This research was supported by the Basic Science Research Program through
NRF of Korea under Grant No. 2012-0002245.
W.E.P was supported by DOE under Grant No.
DE-FG02-04ER46111.

\end{document}